\documentclass[preprint,aps,prd,a4paper,tightenlines,nofootinbib,showpacs]{revtex4-2}
\usepackage{amsfonts,amsmath,amssymb,graphicx,mathrsfs,color}
\usepackage{bm} 
\usepackage{feynmp}
\usepackage{ifpdf}
\ifpdf
\fi
\makeatletter
\def\endfmffile{%
  \fmfcmd{\p@rcent\space the end.^^J%
          end.^^J%
          endinput;}%
  \if@fmfio
    \immediate\closeout\@outfmf
  \fi
  \IfFileExists{\thefmffile.mp}{\immediate\write18{mpost \thefmffile}}{}
  \let\thefmffile\relax
}
\makeatother
\newcommand{\ben}{\begin{displaymath}}
\newcommand{\een}{\end{displaymath}}
\newcommand{\be}{\begin{equation}}
\newcommand{\ee}{\end{equation}}
\newcommand{\bea}{\begin{eqnarray}}
\newcommand{\eea}{\end{eqnarray}}

\newcommand{\q}{\bar{q}}
\newcommand{\D}{\bar{D}}

\newcommand{\bc}{\begin{center}}
\newcommand{\ec}{\end{center}}

\newcommand{\eqn}[1]{\label{#1}}
\newcommand{\eq}[1]{Eq.~(\ref{#1})}
\newcommand{\eqs}[1]{Eqs.~(\ref{#1})}
\newcommand{\fign}[1]{\label{#1}}
\newcommand{\fig}[1]{Fig.~\ref{#1}}

\begin{document}
\title{Comment on ``$\sigma$-meson: Four-quark versus two-quark components and decay width in a Bethe-Salpeter approach"}
\author{B. Blankleider}
\affiliation{
College of Science and Engineering,
 Flinders University, Bedford Park, SA 5042, Australia}
\email{boris.blankleider@flinders.edu.au}
\author{A. N. Kvinikhidze}
\affiliation{Andrea Razmadze Mathematical 
Institute of Tbilisi State University, 6, Tamarashvili Str., 0186 Tbilisi, Georgia}
\email{sasha\_kvinikhidze@hotmail.com}
\date{\today}

\begin{abstract}
In a recent paper by N.\ Santowsky {\it et al.} [Phys.\ Rev.\ D {\bf 102}, 056014 (2020)], covariant coupled equations were derived to describe
a tetraquark in terms of a mix of four-quark states $2q2\q$ and two-quark states $q\q$. These equations were expressed in terms of vertices describing the disintegration of a tetraquark into identical two-meson states, into a diquark-antidiquark pair, and into a quark-antiquark pair. We show that these equations are inconsistent as they imply a $q\q$ Bethe-Salpeter kernel that is $q\q$-reducible.
\end{abstract}
 
\maketitle
\newpage
In 2012, Heupel, Eichmann and Fischer (HEF) \cite{Heupel:2012ua} developed covariant equations describing a tetraquark using a model where the two-quark plus two-antiquark ($2q2\q$) system is described by four-body ($4q$) Faddeev-like equations of Khvedelidze and  Kvinikhidze \cite{Khvedelidze:1991qb}, and where the dynamics is dominated by the formation of either two identical mesons or a diquark-antidiquark pair. These equations are represented graphically in \fig{fig:1}, and relate the form factors $\phi_M$ and $\phi_D$ of the tetraquark, describing its disintegration into two identical mesons, and a diquark-antidiquark pair, respectively. As is evident from \fig{fig:1}, the input interactions to these equations consist of vertices for the transitions between a meson ($M$) and a quark-antiquark pair ($q\q\leftrightarrow M$), a diquark ($D$) and a quark-quark pair ($q q\leftrightarrow D$), and between an antidiquark ($\D$) and an antiquark-antiquark pair ($\q \q\leftrightarrow \bar D$). Missing  from these equations is the phenomenon of quark-antiquark annihilation which would result in coupling to two-body ($2q$) $q\q$ states.

There have since been two attempts to extend the equations of HEF to include coupling to $q\q$ channels. The first of these was our derivation of 2014 \cite{Kvinikhidze:2014yqa} where  disconnected contributions were added to the usual connected part of the  $q\q$ interaction. The second was a recent derivation of Santowsky {\it et al.} (SEFWW)  \cite{Santowsky:2020pwd} where coupling to $q\q$ channels was included phenomenologically. The tetraquark equations of SEFWW are represented graphically in \fig{fig:2}, and include an additional tetraquark form factor $\Gamma^*$, describing the disintegration of a tetraquark into a $q\q$ pair.
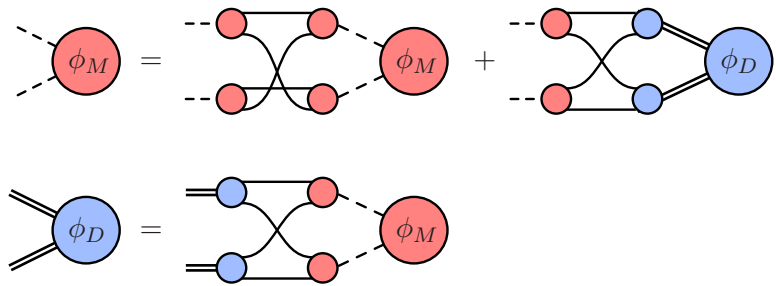
\begin{figure}[b]
\begin{center}
\begin{fmffile}{no2q}
\begin{align*}
\parbox{10mm}{
\begin{fmfgraph*}(10,15)
\fmfstraight
\fmfleftn{l}{7}\fmfrightn{r}{7}\fmfbottomn{b}{6}\fmftopn{t}{6}
\fmf{phantom}{l2,mb2,r2}
\fmf{phantom}{l6,mt2,r6}
\fmffreeze
\fmf{phantom}{r2,phi,r6}
\fmffreeze
\fmf{dashes}{phi,l2}
\fmf{dashes}{phi,l6}
\fmfv{d.sh=circle,d.f=empty,d.si=25,label=$\hspace{-4.5mm}\phi_M$,background=(1,,.51,,.5)}{phi}
\end{fmfgraph*}}
\hspace{6mm} &= \hspace{2mm}
\parbox{30mm}{
\begin{fmfgraph*}(30,15)
\fmfstraight
\fmfleftn{l}{7}\fmfrightn{r}{7}\fmfbottomn{b}{6}\fmftopn{t}{6}
\fmf{phantom}{l2,vb1,mb1,vb2,mb2,r2}
\fmf{phantom}{l6,vt1,mt1,vt2,mt2,r6}
\fmf{phantom}{l2,xb1,mb1,xb2,mb2,r2}
\fmf{phantom}{l6,xt1,mt1,xt2,mt2,r6}
\fmf{phantom}{l2,yb1,mb1,yb2,mb2,r2}
\fmf{phantom}{l6,yt1,mt1,yt2,mt2,r6}
\fmffreeze
\fmf{phantom}{r2,phi,r6}
\fmffreeze
\fmfshift{4 right}{xb1}
\fmfshift{4 down}{xb1}
\fmfshift{4 left}{xt2}
\fmfshift{4 down}{xt2}
\fmfshift{4 right}{xt1}
\fmfshift{4 down}{xt1}
\fmfshift{4 left}{xb2}
\fmfshift{4 down}{xb2}
\fmfshift{4 right}{yb1}
\fmfshift{4 up}{yb1}
\fmfshift{4 left}{yt2}
\fmfshift{4 up}{yt2}
\fmfshift{4 right}{yt1}
\fmfshift{4 up}{yt1}
\fmfshift{4 left}{yb2}
\fmfshift{4 up}{yb2}
\fmfv{d.sh=circle,d.f=empty,d.si=11,background=(1,,.51,,.5)}{vb1}
\fmfv{d.sh=circle,d.f=empty,d.si=11,background=(1,,.51,,.5)}{vb2}
\fmfv{d.sh=circle,d.f=empty,d.si=11,background=(1,,.51,,.5)}{vt1}
\fmfv{d.sh=circle,d.f=empty,d.si=11,background=(1,,.51,,.5)}{vt2}
\fmf{dashes}{l2,vb1}
\fmf{dashes}{l6,vt1}
\fmf{dashes}{phi,xb2}
\fmf{dashes}{phi,yt2}
\fmf{phantom}{vb1,vb2}
\fmf{phantom}{vt1,vt2}
\fmfi{plain}{vloc(__xt2) {left}..tension 1 ..{left}vloc(__xb1)}
\fmfi{plain}{vloc(__xt1) {right}.. tension 1 ..{right}vloc(__xb2)}
\fmfi{plain}{vloc(__yt2) .. vloc(__yt1)}
\fmfi{plain}{vloc(__yb1) .. vloc(__yb2)}
\fmfv{d.sh=circle,d.f=empty,d.si=25,label=$\hspace{-4.5mm}\phi_M$,background=(1,,.51,,.5)}{phi}
\end{fmfgraph*}}
\hspace{7mm}+\hspace{1mm}
\parbox{30mm}{
\begin{fmfgraph*}(30,15)
\fmfstraight
\fmfleftn{l}{7}\fmfrightn{r}{7}\fmfbottomn{b}{6}\fmftopn{t}{6}
\fmf{phantom}{l2,vb1,mb1,vb2,mb2,r2}
\fmf{phantom}{l6,vt1,mt1,vt2,mt2,r6}
\fmf{phantom}{l2,xb1,mb1,xb2,mb2,r2}
\fmf{phantom}{l6,xt1,mt1,xt2,mt2,r6}
\fmf{phantom}{l2,yb1,mb1,yb2,mb2,r2}
\fmf{phantom}{l6,yt1,mt1,yt2,mt2,r6}
\fmffreeze
\fmf{phantom}{r2,phi,r6}
\fmffreeze
\fmfshift{4 right}{xb1}
\fmfshift{4 down}{xb1}
\fmfshift{4 left}{xt2}
\fmfshift{4 down}{xt2}
\fmfshift{4 right}{xt1}
\fmfshift{4 down}{xt1}
\fmfshift{4 left}{xb2}
\fmfshift{4 down}{xb2}
\fmfshift{4 right}{yb1}
\fmfshift{4 up}{yb1}
\fmfshift{4 left}{yt2}
\fmfshift{4 up}{yt2}
\fmfshift{4 right}{yt1}
\fmfshift{4 up}{yt1}
\fmfshift{4 left}{yb2}
\fmfshift{4 up}{yb2}
\fmfv{d.sh=circle,d.f=empty,d.si=11,background=(1,,.51,,.5)}{vb1}
\fmfv{d.sh=circle,d.f=empty,d.si=11,background=(.6235,,.7412,,1)}{vb2}
\fmfv{d.sh=circle,d.f=empty,d.si=11,background=(1,,.51,,.5)}{vt1}
\fmfv{d.sh=circle,d.f=empty,d.si=11,background=(.6235,,.7412,,1)}{vt2}
\fmf{dashes}{l2,vb1}
\fmf{dashes}{l6,vt1}
\fmf{dbl_plain}{phi,xb2}
\fmf{dbl_plain}{phi,yt2}
\fmf{phantom}{vb1,vb2}
\fmf{phantom}{vt1,vt2}
\fmfi{plain}{vloc(__xt2) {left}..tension 1 ..{left}vloc(__yb1)}
\fmfi{plain}{vloc(__xt1) {right}.. tension 1 ..{right}vloc(__yb2)}
\fmfi{plain}{vloc(__yt2) .. vloc(__yt1)}
\fmfi{plain}{vloc(__xb1) .. vloc(__xb2)}
\fmfv{d.sh=circle,d.f=empty,d.si=25,label=$\hspace{-4.5mm}\phi_D$,background=(.6235,,.7412,,1)}{phi}
\end{fmfgraph*}}\\[6mm]
\parbox{10mm}{
\begin{fmfgraph*}(10,15)
\fmfstraight
\fmfleftn{l}{7}\fmfrightn{r}{7}\fmfbottomn{b}{6}\fmftopn{t}{6}
\fmf{phantom}{l2,mb2,r2}
\fmf{phantom}{l6,mt2,r6}
\fmffreeze
\fmf{phantom}{r2,phi,r6}
\fmffreeze
\fmf{dbl_plain}{phi,l2}
\fmf{dbl_plain}{phi,l6}
\fmfv{d.sh=circle,d.f=empty,d.si=25,label=$\hspace{-4.5mm}\phi_D$,background=(.6235,,.7412,,1)}{phi}
\end{fmfgraph*}}
\hspace{6mm} &= \hspace{2mm}
\parbox{30mm}{
\begin{fmfgraph*}(30,15)
\fmfstraight
\fmfleftn{l}{7}\fmfrightn{r}{7}\fmfbottomn{b}{6}\fmftopn{t}{6}
\fmf{phantom}{l2,vb1,mb1,vb2,mb2,r2}
\fmf{phantom}{l6,vt1,mt1,vt2,mt2,r6}
\fmf{phantom}{l2,xb1,mb1,xb2,mb2,r2}
\fmf{phantom}{l6,xt1,mt1,xt2,mt2,r6}
\fmf{phantom}{l2,yb1,mb1,yb2,mb2,r2}
\fmf{phantom}{l6,yt1,mt1,yt2,mt2,r6}
\fmffreeze
\fmf{phantom}{r2,phi,r6}
\fmffreeze
\fmfshift{4 right}{xb1}
\fmfshift{4 down}{xb1}
\fmfshift{4 left}{xt2}
\fmfshift{4 down}{xt2}
\fmfshift{4 right}{xt1}
\fmfshift{4 down}{xt1}
\fmfshift{4 left}{xb2}
\fmfshift{4 down}{xb2}
\fmfshift{4 right}{yb1}
\fmfshift{4 up}{yb1}
\fmfshift{4 left}{yt2}
\fmfshift{4 up}{yt2}
\fmfshift{4 right}{yt1}
\fmfshift{4 up}{yt1}
\fmfshift{4 left}{yb2}
\fmfshift{4 up}{yb2}
\fmfv{d.sh=circle,d.f=empty,d.si=11,background=(.6235,,.7412,,1)}{vb1}
\fmfv{d.sh=circle,d.f=empty,d.si=11,background=(1,,.51,,.5)}{vb2}
\fmfv{d.sh=circle,d.f=empty,d.si=11,background=(.6235,,.7412,,1)}{vt1}
\fmfv{d.sh=circle,d.f=empty,d.si=11,background=(1,,.51,,.5)}{vt2}
\fmf{dbl_plain}{l2,vb1}
\fmf{dbl_plain}{l6,vt1}
\fmf{dashes}{phi,xb2}
\fmf{dashes}{phi,yt2}
\fmf{phantom}{vb1,vb2}
\fmf{phantom}{vt1,vt2}
\fmfi{plain}{vloc(__xt2) {left}..tension 1 ..{left}vloc(__yb1)}
\fmfi{plain}{vloc(__xt1) {right}.. tension 1 ..{right}vloc(__yb2)}
\fmfi{plain}{vloc(__yt2) .. vloc(__yt1)}
\fmfi{plain}{vloc(__xb1) .. vloc(__xb2)}
\fmfv{d.sh=circle,d.f=empty,d.si=25,label=$\hspace{-4.5mm}\phi_M$,background=(1,,.51,,.5)}{phi}
\end{fmfgraph*}}
\end{align*}
\end{fmffile}   
\vspace{1mm}

\caption{\fign{fig:1}  Tetraquark equations without coupling to $q\q$ channels, as first developed in Ref.\ \cite{Heupel:2012ua}. Tetraquark form factors $\phi_M$ (displayed in red) couple to two mesons (dashed lines), and tetraquark form factors $\phi_D$ (displayed in blue) couple to diquark and antidiquark states (double-lines).}
\end{center}
\end{figure}
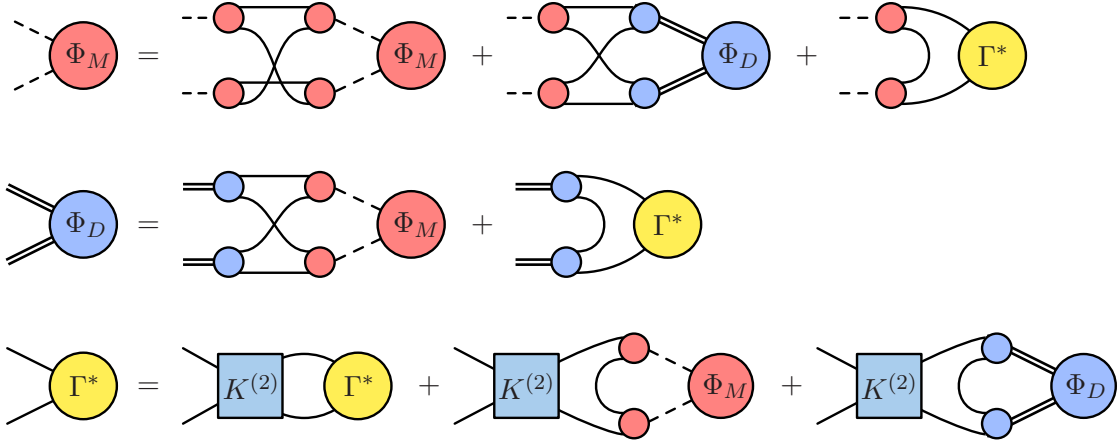
\begin{figure}[t]
\begin{center}
\begin{fmffile}{yes2q}
\begin{align*}
\parbox{10mm}{
\begin{fmfgraph*}(10,15)
\fmfstraight
\fmfleftn{l}{7}\fmfrightn{r}{7}\fmfbottomn{b}{6}\fmftopn{t}{6}
\fmf{phantom}{l2,mb2,r2}
\fmf{phantom}{l6,mt2,r6}
\fmffreeze
\fmf{phantom}{r2,phi,r6}
\fmffreeze
\fmf{dashes}{phi,l2}
\fmf{dashes}{phi,l6}
\fmfv{d.sh=circle,d.f=empty,d.si=25,label=$\hspace{-4.5mm}\Phi_M$,background=(1,,.51,,.5)}{phi}
\end{fmfgraph*}}
\hspace{6mm} &=\hspace{2mm}
\parbox{30mm}{
\begin{fmfgraph*}(30,15)
\fmfstraight
\fmfleftn{l}{7}\fmfrightn{r}{7}\fmfbottomn{b}{6}\fmftopn{t}{6}
\fmf{phantom}{l2,vb1,mb1,vb2,mb2,r2}
\fmf{phantom}{l6,vt1,mt1,vt2,mt2,r6}
\fmf{phantom}{l2,xb1,mb1,xb2,mb2,r2}
\fmf{phantom}{l6,xt1,mt1,xt2,mt2,r6}
\fmf{phantom}{l2,yb1,mb1,yb2,mb2,r2}
\fmf{phantom}{l6,yt1,mt1,yt2,mt2,r6}
\fmffreeze
\fmf{phantom}{r2,phi,r6}
\fmffreeze
\fmfshift{4 right}{xb1}
\fmfshift{4 down}{xb1}
\fmfshift{4 left}{xt2}
\fmfshift{4 down}{xt2}
\fmfshift{4 right}{xt1}
\fmfshift{4 down}{xt1}
\fmfshift{4 left}{xb2}
\fmfshift{4 down}{xb2}
\fmfshift{4 right}{yb1}
\fmfshift{4 up}{yb1}
\fmfshift{4 left}{yt2}
\fmfshift{4 up}{yt2}
\fmfshift{4 right}{yt1}
\fmfshift{4 up}{yt1}
\fmfshift{4 left}{yb2}
\fmfshift{4 up}{yb2}
\fmfv{d.sh=circle,d.f=empty,d.si=11,background=(1,,.51,,.5)}{vb1}
\fmfv{d.sh=circle,d.f=empty,d.si=11,background=(1,,.51,,.5)}{vb2}
\fmfv{d.sh=circle,d.f=empty,d.si=11,background=(1,,.51,,.5)}{vt1}
\fmfv{d.sh=circle,d.f=empty,d.si=11,background=(1,,.51,,.5)}{vt2}
\fmf{dashes}{l2,vb1}
\fmf{dashes}{l6,vt1}
\fmf{dashes}{phi,xb2}
\fmf{dashes}{phi,yt2}
\fmf{phantom}{vb1,vb2}
\fmf{phantom}{vt1,vt2}
\fmfi{plain}{vloc(__xt2) {left}..tension 1 ..{left}vloc(__xb1)}
\fmfi{plain}{vloc(__xt1) {right}.. tension 1 ..{right}vloc(__xb2)}
\fmfi{plain}{vloc(__yt2) .. vloc(__yt1)}
\fmfi{plain}{vloc(__yb1) .. vloc(__yb2)}
\fmfv{d.sh=circle,d.f=empty,d.si=25,label=$\hspace{-4.5mm}\Phi_M$,background=(1,,.51,,.5)}{phi}
\end{fmfgraph*}}
\hspace{7mm}+\hspace{1mm}
\parbox{30mm}{
\begin{fmfgraph*}(30,15)
\fmfstraight
\fmfleftn{l}{7}\fmfrightn{r}{7}\fmfbottomn{b}{6}\fmftopn{t}{6}
\fmf{phantom}{l2,vb1,mb1,vb2,mb2,r2}
\fmf{phantom}{l6,vt1,mt1,vt2,mt2,r6}
\fmf{phantom}{l2,xb1,mb1,xb2,mb2,r2}
\fmf{phantom}{l6,xt1,mt1,xt2,mt2,r6}
\fmf{phantom}{l2,yb1,mb1,yb2,mb2,r2}
\fmf{phantom}{l6,yt1,mt1,yt2,mt2,r6}
\fmffreeze
\fmf{phantom}{r2,phi,r6}
\fmffreeze
\fmfshift{4 right}{xb1}
\fmfshift{4 down}{xb1}
\fmfshift{4 left}{xt2}
\fmfshift{4 down}{xt2}
\fmfshift{4 right}{xt1}
\fmfshift{4 down}{xt1}
\fmfshift{4 left}{xb2}
\fmfshift{4 down}{xb2}
\fmfshift{4 right}{yb1}
\fmfshift{4 up}{yb1}
\fmfshift{4 left}{yt2}
\fmfshift{4 up}{yt2}
\fmfshift{4 right}{yt1}
\fmfshift{4 up}{yt1}
\fmfshift{4 left}{yb2}
\fmfshift{4 up}{yb2}
\fmfv{d.sh=circle,d.f=empty,d.si=11,background=(1,,.51,,.5)}{vb1}
\fmfv{d.sh=circle,d.f=empty,d.si=11,background=(.6235,,.7412,,1)}{vb2}
\fmfv{d.sh=circle,d.f=empty,d.si=11,background=(1,,.51,,.5)}{vt1}
\fmfv{d.sh=circle,d.f=empty,d.si=11,background=(.6235,,.7412,,1)}{vt2}
\fmf{dashes}{l2,vb1}
\fmf{dashes}{l6,vt1}
\fmf{dbl_plain}{phi,xb2}
\fmf{dbl_plain}{phi,yt2}
\fmf{phantom}{vb1,vb2}
\fmf{phantom}{vt1,vt2}
\fmfi{plain}{vloc(__xt2) {left}..tension 1 ..{left}vloc(__yb1)}
\fmfi{plain}{vloc(__xt1) {right}.. tension 1 ..{right}vloc(__yb2)}
\fmfi{plain}{vloc(__yt2) .. vloc(__yt1)}
\fmfi{plain}{vloc(__xb1) .. vloc(__xb2)}
\fmfv{d.sh=circle,d.f=empty,d.si=25,label=$\hspace{-4.5mm}\Phi_D$,background=(.6235,,.7412,,1)}{phi}
\end{fmfgraph*}}
\hspace{7mm}+\hspace{2mm}
\parbox{20mm}{
\begin{fmfgraph*}(20,15)
\fmfstraight
\fmfleftn{l}{7}\fmfrightn{r}{7}\fmfbottomn{b}{4}\fmftopn{t}{4}
\fmf{phantom}{l2,vb1,mb1,r2}
\fmf{phantom}{l6,vt1,mt1,r6}
\fmf{phantom}{l2,xb1,mb1,r2}
\fmf{phantom}{l6,xt1,mt1,r6}
\fmf{phantom}{l2,yb1,mb1,r2}
\fmf{phantom}{l6,yt1,mt1,r6}
\fmffreeze
\fmf{phantom}{r2,phi,r6}
\fmffreeze
\fmfshift{4 right}{xb1}
\fmfshift{4 down}{xb1}
\fmfshift{4 right}{xt1}
\fmfshift{4 down}{xt1}
\fmfshift{4 right}{yb1}
\fmfshift{4 up}{yb1}
\fmfshift{4 right}{yt1}
\fmfshift{4 up}{yt1}
\fmfv{d.sh=circle,d.f=empty,d.si=11,background=(1,,.51,,.5)}{vb1}
\fmfv{d.sh=circle,d.f=empty,d.si=11,background=(1,,.51,,.5)}{vt1}
\fmf{dashes}{l2,vb1}
\fmf{dashes}{l6,vt1}
\fmfi{plain}{vloc(__yb1) {right}.. tension 1 ..{left}vloc(__xt1)}
\fmfi{plain}{vloc(__yt1) {right}.. vloc(__phi)}
\fmfi{plain}{vloc(__xb1){right} .. vloc(__phi)}
\fmfv{d.sh=circle,d.f=empty,d.si=25,label=$\hspace{-4.mm}\Gamma^*$,background=(1,,.9333,,.333)}{phi}
\end{fmfgraph*}}\\[6mm]
\parbox{10mm}{
\begin{fmfgraph*}(10,15)
\fmfstraight
\fmfleftn{l}{7}\fmfrightn{r}{7}\fmfbottomn{b}{6}\fmftopn{t}{6}
\fmf{phantom}{l2,mb2,r2}
\fmf{phantom}{l6,mt2,r6}
\fmffreeze
\fmf{phantom}{r2,phi,r6}
\fmffreeze
\fmf{dbl_plain}{phi,l2}
\fmf{dbl_plain}{phi,l6}
\fmfv{d.sh=circle,d.f=empty,d.si=25,label=$\hspace{-4.5mm}\Phi_D$,background=(.6235,,.7412,,1)}{phi}
\end{fmfgraph*}}
\hspace{6mm} &= \hspace{2mm}
\parbox{30mm}{
\begin{fmfgraph*}(30,15)
\fmfstraight
\fmfleftn{l}{7}\fmfrightn{r}{7}\fmfbottomn{b}{6}\fmftopn{t}{6}
\fmf{phantom}{l2,vb1,mb1,vb2,mb2,r2}
\fmf{phantom}{l6,vt1,mt1,vt2,mt2,r6}
\fmf{phantom}{l2,xb1,mb1,xb2,mb2,r2}
\fmf{phantom}{l6,xt1,mt1,xt2,mt2,r6}
\fmf{phantom}{l2,yb1,mb1,yb2,mb2,r2}
\fmf{phantom}{l6,yt1,mt1,yt2,mt2,r6}
\fmffreeze
\fmf{phantom}{r2,phi,r6}
\fmffreeze
\fmfshift{4 right}{xb1}
\fmfshift{4 down}{xb1}
\fmfshift{4 left}{xt2}
\fmfshift{4 down}{xt2}
\fmfshift{4 right}{xt1}
\fmfshift{4 down}{xt1}
\fmfshift{4 left}{xb2}
\fmfshift{4 down}{xb2}
\fmfshift{4 right}{yb1}
\fmfshift{4 up}{yb1}
\fmfshift{4 left}{yt2}
\fmfshift{4 up}{yt2}
\fmfshift{4 right}{yt1}
\fmfshift{4 up}{yt1}
\fmfshift{4 left}{yb2}
\fmfshift{4 up}{yb2}
\fmfv{d.sh=circle,d.f=empty,d.si=11,background=(.6235,,.7412,,1)}{vb1}
\fmfv{d.sh=circle,d.f=empty,d.si=11,background=(1,,.51,,.5)}{vb2}
\fmfv{d.sh=circle,d.f=empty,d.si=11,background=(.6235,,.7412,,1)}{vt1}
\fmfv{d.sh=circle,d.f=empty,d.si=11,background=(1,,.51,,.5)}{vt2}
\fmf{dbl_plain}{l2,vb1}
\fmf{dbl_plain}{l6,vt1}
\fmf{dashes}{phi,xb2}
\fmf{dashes}{phi,yt2}
\fmf{phantom}{vb1,vb2}
\fmf{phantom}{vt1,vt2}
\fmfi{plain}{vloc(__xt2) {left}..tension 1 ..{left}vloc(__yb1)}
\fmfi{plain}{vloc(__xt1) {right}.. tension 1 ..{right}vloc(__yb2)}
\fmfi{plain}{vloc(__yt2) .. vloc(__yt1)}
\fmfi{plain}{vloc(__xb1) .. vloc(__xb2)}
\fmfv{d.sh=circle,d.f=empty,d.si=25,label=$\hspace{-4.5mm}\Phi_M$,background=(1,,.51,,.5)}{phi}
\end{fmfgraph*}}
\hspace{7mm}+\hspace{2mm}
\parbox{20mm}{
\begin{fmfgraph*}(20,15)
\fmfstraight
\fmfleftn{l}{7}\fmfrightn{r}{7}\fmfbottomn{b}{4}\fmftopn{t}{4}
\fmf{phantom}{l2,vb1,mb1,r2}
\fmf{phantom}{l6,vt1,mt1,r6}
\fmf{phantom}{l2,xb1,mb1,r2}
\fmf{phantom}{l6,xt1,mt1,r6}
\fmf{phantom}{l2,yb1,mb1,r2}
\fmf{phantom}{l6,yt1,mt1,r6}
\fmffreeze
\fmf{phantom}{r2,phi,r6}
\fmffreeze
\fmfshift{4 right}{xb1}
\fmfshift{4 down}{xb1}
\fmfshift{4 right}{xt1}
\fmfshift{4 down}{xt1}
\fmfshift{4 right}{yb1}
\fmfshift{4 up}{yb1}
\fmfshift{4 right}{yt1}
\fmfshift{4 up}{yt1}
\fmfv{d.sh=circle,d.f=empty,d.si=11,background=(.6235,,.7412,,1)}{vb1}
\fmfv{d.sh=circle,d.f=empty,d.si=11,background=(.6235,,.7412,,1)}{vt1}
\fmf{dbl_plain}{l2,vb1}
\fmf{dbl_plain}{l6,vt1}
\fmfi{plain}{vloc(__yb1) {right}.. tension 1 ..{left}vloc(__xt1)}
\fmfi{plain}{vloc(__yt1) {right}.. vloc(__phi)}
\fmfi{plain}{vloc(__xb1){right} .. vloc(__phi)}
\fmfv{d.sh=circle,d.f=empty,d.si=25,label=$\hspace{-4.mm}\Gamma^*$,background=(1,,.9333,,.333)}{phi}
\end{fmfgraph*}}\\[5mm]
\parbox{10mm}{
\begin{fmfgraph*}(10,15)
\fmfstraight
\fmfleftn{l}{7}\fmfrightn{r}{7}\fmfbottomn{b}{3}\fmftopn{t}{3}
\fmf{phantom}{l2,mb2,r2}
\fmf{phantom}{l6,mt2,r6}
\fmffreeze
\fmf{phantom}{r2,phi,r6}
\fmffreeze
\fmf{plain}{phi,l2}
\fmf{plain}{phi,l6}
\fmfv{d.sh=circle,d.f=empty,d.si=25,label=$\hspace{-4.mm}\Gamma^*$,background=(1,,.9333,,.333)}{phi}
\end{fmfgraph*}}
\hspace{6mm} &=\hspace{2mm}
\parbox{23mm}{
\begin{fmfgraph*}(23,15)
\fmfstraight
\fmfleftn{l}{7}\fmfrightn{r}{7}\fmfbottomn{b}{5}\fmftopn{t}{5}
\fmf{phantom}{l2,mb1,vb1,mb2,r2}
\fmf{phantom}{l6,mt1,vt1,mt2,r6}
\fmf{phantom,tension=.3}{l2,vb1}
\fmf{phantom,tension=.3}{l6,vt1}
\fmffreeze
\fmf{phantom}{r2,phi,r6}
\fmffreeze
\fmf{phantom}{vb1,delta,vt1}
\fmffreeze
\fmf{plain}{l2,delta}
\fmf{plain}{l6,delta}
\fmf{plain,left=.6}{delta,phi}
\fmf{plain,right=.6}{delta,phi}
\fmfv{d.sh=circle,d.f=empty,d.si=25,label=$\hspace{-4.mm}\Gamma^*$,background=(1,,.9333,,.333)}{phi}
\fmfv{d.sh=square,d.f=empty,d.si=24,background=(0.664,,.805,,.930)}{delta}
\fmfiv{l=$K^{(2)}\hspace{5mm}$,l.a=180,l.d=.0000w}{c}
\end{fmfgraph*}}
\hspace{7mm}+\hspace{1mm}
\parbox{35mm}{
\begin{fmfgraph*}(35,15)
\fmfstraight
\fmfleftn{l}{7}\fmfrightn{r}{7}\fmfbottomn{b}{7}\fmftopn{t}{7}
\fmf{phantom}{l2,mb1,vb1,mb2,vb2,mb3,r2}
\fmf{phantom}{l6,mt1,vt1,mt2,vt2,mt3,r6}
\fmf{phantom}{l2,mb1,xb1,mb2,xb2,mb3,r2}
\fmf{phantom}{l6,mt1,xt1,mt2,xt2,mt3,r6}
\fmf{phantom}{l2,mb1,yb1,mb2,yb2,mb3,r2}
\fmf{phantom}{l6,mt1,yt1,mt2,yt2,mt3,r6}
\fmf{phantom,tension=6}{l2,mb1}
\fmf{phantom,tension=6}{l6,mt1}
\fmf{phantom,tension=.12}{r2,vb2}
\fmf{phantom,tension=.12}{r6,vt2}
\fmf{phantom,tension=.12}{r2,xb2}
\fmf{phantom,tension=.12}{r6,xt2}
\fmf{phantom,tension=.12}{r2,yb2}
\fmf{phantom,tension=.12}{r6,yt2}
\fmffreeze
\fmf{phantom}{mb2,c2,mt2}
\fmf{phantom}{r2,phi,r6}
\fmf{phantom}{vb1,delta,vt1}
\fmfv{d.sh=square,d.f=empty,d.si=24,background=(0.664,,.805,,.930)}{delta}
\fmfiv{l=$K^{(2)}\hspace{16mm}$,l.a=180,l.d=.0000w}{c}
\fmffreeze
\fmfshift{9 left}{c2}
\fmfshift{4 left}{xt2}
\fmfshift{4 down}{xt2}
\fmfshift{4 left}{xb2}
\fmfshift{4 down}{xb2}
\fmfshift{4 left}{yt2}
\fmfshift{4 up}{yt2}
\fmfshift{4 left}{yb2}
\fmfshift{4 up}{yb2}
\fmfv{d.sh=circle,d.f=empty,d.si=11,background=(1,,.51,,.5)}{vb2}
\fmfv{d.sh=circle,d.f=empty,d.si=11,background=(1,,.51,,.5)}{vt2}
\fmf{plain}{l2,delta}
\fmf{plain}{l6,delta}
\fmf{dashes}{phi,xb2}
\fmf{dashes}{phi,yt2}
\fmfi{plain}{vloc(__xt2) {left}..tension 1 ..{right}vloc(__yb2)}
\fmfi{plain}{vloc(__yt2) {left}..tension 3 ..{down}vloc(__delta)}
\fmfi{plain}{vloc(__xb2) {left}..tension 3 ..{up}vloc(__delta)}
\fmfv{d.sh=circle,d.f=empty,d.si=24,label=$\hspace{-4.5mm}\Phi_M$,background=(1,,.51,,.5)}{phi}
\end{fmfgraph*}}
\hspace{7mm}+\hspace{1mm}
\parbox{35mm}{
\begin{fmfgraph*}(35,15)
\fmfstraight
\fmfleftn{l}{7}\fmfrightn{r}{7}\fmfbottomn{b}{7}\fmftopn{t}{7}
\fmf{phantom}{l2,mb1,vb1,mb2,vb2,mb3,r2}
\fmf{phantom}{l6,mt1,vt1,mt2,vt2,mt3,r6}
\fmf{phantom}{l2,mb1,xb1,mb2,xb2,mb3,r2}
\fmf{phantom}{l6,mt1,xt1,mt2,xt2,mt3,r6}
\fmf{phantom}{l2,mb1,yb1,mb2,yb2,mb3,r2}
\fmf{phantom}{l6,mt1,yt1,mt2,yt2,mt3,r6}
\fmf{phantom,tension=6}{l2,mb1}
\fmf{phantom,tension=6}{l6,mt1}
\fmf{phantom,tension=.12}{r2,vb2}
\fmf{phantom,tension=.12}{r6,vt2}
\fmf{phantom,tension=.12}{r2,xb2}
\fmf{phantom,tension=.12}{r6,xt2}
\fmf{phantom,tension=.12}{r2,yb2}
\fmf{phantom,tension=.12}{r6,yt2}
\fmffreeze
\fmf{phantom}{mb2,c2,mt2}
\fmf{phantom}{r2,phi,r6}
\fmf{phantom}{vb1,delta,vt1}
\fmfv{d.sh=square,d.f=empty,d.si=24,background=(0.664,,.805,,.930)}{delta}
\fmfiv{l=$K^{(2)}\hspace{16mm}$,l.a=180,l.d=.0000w}{c}
\fmffreeze
\fmfshift{9 left}{c2}
\fmfshift{4 left}{xt2}
\fmfshift{4 down}{xt2}
\fmfshift{4 left}{xb2}
\fmfshift{4 down}{xb2}
\fmfshift{4 left}{yt2}
\fmfshift{4 up}{yt2}
\fmfshift{4 left}{yb2}
\fmfshift{4 up}{yb2}
\fmfv{d.sh=circle,d.f=empty,d.si=11,background=(.6235,,.7412,,1)}{vb2}
\fmfv{d.sh=circle,d.f=empty,d.si=11,background=(.6235,,.7412,,1)}{vt2}
\fmf{plain}{l2,delta}
\fmf{plain}{l6,delta}
\fmf{dbl_plain}{phi,xb2}
\fmf{dbl_plain}{phi,yt2}
\fmfi{plain}{vloc(__xt2) {left}..tension 1 ..{right}vloc(__yb2)}
\fmfi{plain}{vloc(__yt2) {left}..tension 3 ..{down}vloc(__delta)}
\fmfi{plain}{vloc(__xb2) {left}..tension 3 ..{up}vloc(__delta)}
\fmfv{d.sh=circle,d.f=empty,d.si=24,label=$\hspace{-4.5mm}\Phi_D$,background=(.6235,,.7412,,1)}{phi}
\end{fmfgraph*}}
\end{align*}
\end{fmffile}   
\vspace{3mm}

\caption{\fign{fig:2} The tetraquark equations of SEFWW \cite{Santowsky:2020pwd} which include coupling to $q\q$ channels. In addition to the tetraquark form factors as in \fig{fig:1}, these equations involve the tetraquark form factor $\Gamma^*$ (displayed in yellow) that couples to $q\q$ states (solid lines). The amplitude $K^{(2)}$ (displayed in light blue) represents the $q\q$ kernel in a theory without $q\q$ annihilation.}
\end{center}
\end{figure}

For the tetraquark equations of \fig{fig:2} to be meaningful, it is essential that the form factor $\Gamma^*$ be identified with the residue of the $q\q$ Green function $G^{(2)}$ describing the formation of the tetraquark in the scattering of a quark from an antiquark; that is, as $P^2 \rightarrow M^2$, where $P$ is the total momentum of the $q\q$ system and $M$ is the mass of the tetraquark, 
\be
G^{(2)}\rightarrow \frac{G_0^{(2)}\Gamma^* \bar\Gamma^*G_0^{(2)}}{P^2-M^2},
\ee
where $G_0^{(2)}$ is the fully disconnected $q\q$ propagator corresponding to the independent propagation of $q$ and $\q$ in the $s$ channel. This implies that $\Gamma^*$ satisfies the bound state equation
\be
\Gamma^* = K_{ir} G_0^{(2)}\Gamma^*  \eqn{bs}
\ee
where $K_{ir}$ is the $q\q$-irreducible Bethe-Salpeter kernel for the $q\q$ system.

Here we would like to point out that the tetraquark equations of SEFWW cannot be correct as they imply a kernel $K_{ir}$ that is $q\q$-reducible. To show this, we write the three coupled equations corresponding to \fig{fig:2} as\footnote{For simplicity, we ignore all symmetry factors in \eqs{SEFWW} as they do not affect our argument.}
\begin{subequations}  \eqn{SEFWW}
\begin{align}
\Phi_M&=
V_{MM} G^0_{M} \Phi_M  +V_{MD} G^0_{D} \Phi_D +  N_M G_0^{(2)}\Gamma^*, \\[2mm]
\Phi_D &=V_{DM} G^0_{M} \Phi_M  +  N_D G_0^{(2)}\Gamma^*, \\[2mm]
\Gamma^*&=K^{(2)} G_0^{(2)}\Gamma^*+K^{(2)} G_0^{(2)} \bar N_M G^0_M \Phi_M+K^{(2)} G_0^{(2)} \bar N_D G^0_D \Phi_D,
\end{align}
\end{subequations}
where $V_{MM}$, $V_{MD}$, and $V_{DM}$, are quark-exchange potentials for the processes $MM\leftarrow MM$,   $MM\leftarrow D\D$,  and $D\D\leftarrow MM$,  respectively, and where $N_M$, $N_D$, $\bar N_M$, and $\bar N_D$ describe the transitions between $4q$ and $2q$ states via the processes  $MM\leftarrow q\q$, $D\D\leftarrow q\q$, $q\q\leftarrow MM$, and  $q\q\leftarrow D\D$. Note that these equations also involve a $q\q$ kernel $K^{(2)}$ which should not be confused with $K_{ir}$ as it does not contain terms that involve $2q\leftrightarrow 4q$ transitions. 
 
Writing \eqs{SEFWW} in matrix form as
\begin{subequations}
\begin{align}
\Phi &= V G^0\Phi + N G_0^{(2)}\Gamma^*, \eqn{a}  \\
\Gamma^* & = K^{(2)} G_0^{(2)}\left( \Gamma^* + \bar N G^0 \Phi\right),  \eqn{b}
\end{align}
\end{subequations}
where
\begin{align}
\Phi &= 
\begin{pmatrix}
\Phi_M \\ \Phi_D 
\end{pmatrix}
,\hspace{1cm}
G^0=
\begin{pmatrix}
G^0_M& 0 \\ 
0 & G^0_D
\end{pmatrix},\\[3mm]
N &= 
\begin{pmatrix}
N_M \\ N_D 
\end{pmatrix},\hspace{1cm}
\bar N = 
\begin{pmatrix}
\bar N_M & \bar N_D 
\end{pmatrix},
\end{align}
and
\be
V = \begin{pmatrix}
V_{MM} & V_{MD}\\
V_{DM} & 0 
\end{pmatrix},  \eqn{Vmat}
\ee
one can solve \eq{a} for $\Phi$ and substitute the result into \eq{b} to obtain
\be
\Gamma^* = \left[K^{(2)} + K^{(2)} G_0^{(2)} \bar N G^0 \left( 1-V G^0\right)^{-1} N\right] G_0^{(2)} \Gamma^*.
\ee
Comparison with \eq{bs} shows that
\be
K_{ir} = K^{(2)} + K^{(2)} G_0^{(2)} \bar N G^0 \left( 1-V G^0\right)^{-1} N,
\ee
which is in conflict with the very definition of a $q\q$ kernel since this expression for $K_{ir}$ is $q\q$ reducible (notice the presence of the $q\q$ propagator $G_0^{(2)}$). For this reason the tetraquark equations of Ref.\ \cite{Santowsky:2020pwd} are inconsistent.

\bibliography{/Users/phbb/Physics/Papers/refn}

\end{document}